\documentclass[prb,preprint,superscriptaddress]{revtex4}
\usepackage{graphicx}
\begin{document}

\title{The electrical properties of Cd$_2$Re$_2$O$_7$ under pressure}
\author{N. Bari\v{s}i\'c} \affiliation{IPMC, Facult\'e Science de Base,
EPFL, CH-1015 Lausanne, Switzerland}

\author{L. Forr\'o}

\affiliation{IPMC, Facult\'e Science de Base, EPFL, CH-1015
Lausanne, Switzerland}

\author{D. Mandrus}

\affiliation{Condensed Matter Sciences Division, Oak Ridge
National Laboratory, Oak Ridge, TN 37831, USA}

\author{R. Jin}\affiliation{Department of Physics and Astronomy, The University
of Tennessee, Knoxville, TN 37996, USA}

\author{J. He}

\affiliation{Department of Physics and Astronomy, The University
of Tennessee, Knoxville, TN 37996, USA}

\author{P. Fazekas} \affiliation{Research Institute for Solid State Physics
  and Optics, Budapest 114, P.O.B. 49, H-1525 Hungary}\affiliation{IPMC,
Facult\'e Science de Base, EPFL, CH-1015 Lausanne, Switzerland}

\begin{abstract}
We examine the resistivity and thermopower of single crystal specimens
of the pyrochlore oxide Cd$_2$Re$_2$O$_7$ at pressures up to 2GPa. 
Thermopower proves to be a
sensitive tool in the study of the phase diagram of Cd$_2$Re$_2$O$_7$. The
200K metal-to-metal phase transition is accompanied by a strong increase of
the absolute value of the thermopower. A weaker anomaly allows to identify
a second phase transition at 125K. Following the temperature dependence of
this anomaly, we obtain the corresponding phase boundary up to 1.2GPa, and
argue that it must drop to $T=0$ before $p$ reaches 1.8GPa.
There is a wide temperature range where the electrical properties are
fairly sensitive to pressure, indicating the strong coupling of the
electronic degrees of freedom to the lattice.
\end{abstract}

\maketitle

\section{Introduction}

$3d$ transition metal compounds show a variety of Mott phenomena \cite{Mott},
what is readily understood since for these systems, the electron--electron
interaction (measured by an effective Hubbard $U$) is comparable to the
bandwidth $W$. In contrast, $4d$ and, quite in particular, $5d$ transition
metal compounds are supposed to be less correlated because of their relatively
wide $d$-bands. For example, the $5d$ oxide ReO$_3$ is a good metal with
wide $d$-bands.

There are, however, indications that a few $5d$ compounds also show correlation
effects -- or at least, they show cooperative behavior with no obvious
interpretation in terms of independent electron theory. The Mott localization
aspects of the low-temperature behavior of 1T-TaS$_2$ are long known
\cite{tas2}. Recently, the interest turned towards
pyrochlore oxides \cite{footnote1}. 
Y$_2$Ir$_2$O$_7$ is characterized as a Mott insulator
\cite{Irletter}. Cd$_2$Os$_2$O$_7$ undergoes a metal--insulator
transition at 220K. The phenomenon is remarkably well described as
a BCS-type mean field transition which clearly involves the
ordering of an electronic degree of freedom on the background of a
rigid lattice \cite{Osbig,pureslater}. This led to the proposal of
a Slater transition of the spins \cite{Osbig}, which is further
supported by evidence that charge ordering is unimportant
\cite{pureslater}. Let us note, however, that the supposed
spin-density-like order parameter is experimentally not yet
identified, and it would be no simple matter to postulate it
because of the frustrated nature of the pyrochlore
lattice \cite{footnote2}. Furthermore, whatever the
nature of the order parameter, there is a basic difficulty:
according to band theory, the Fermi surface of Cd$_2$Os$_2$O$_7$
does not seem to be nested \cite{band}, and it is difficult to see
how a small-amplitude order can immediately open gaps all over
the Fermi surface. This difficulty alone suffices to infer that
correlation effects are important.

We may generally ask what is so particular about the above cited compounds
that we may consider them as correlated $5d$ electron systems. The common
feature is the geometrical frustration of the transition metal sublattice
(triangular lattice for 1T-TaS$_2$, pyrochlore for Cd$_2$Os$_2$O$_7$).
The concept of frustration was introduced for Ising antiferromagnets
\cite{toulouse}, and it can be generalized to quantum magnets and
Mott-localized spin-orbital systems with ``antiferro'' interactions.
Generally speaking, frustration prevents two-sublattice order, forcing the
system to choose either multi-sublattice (non-collinear) order, or spin
or/and orbital liquid states. It is less obvious what frustration should
mean for itinerant systems, but one can argue that it involves the
suppression of kinetic energy by desctructive interference, and the
resulting enhanced importance of interaction effects. Weak coupling density
wave phases tend to be eliminated and instead, local correlation effects
like mass enhancement and the opening of a spin gap dominate
\cite{fuji,fluct_exch}.

A clear-cut example of the effect of frustration on weakly interacting
electron systems is provided by the $s$-orbital Hubbard model on the
pyrochlore lattice. The tight binding band structure contains a
twofold degenerate zero-dispersion subband, which gives rise to marked
correlation effects even in the weak coupling limit \cite{isoda}. However,
flat subbands are not expected to occur in real systems. Re and Os
pyrochlore oxides are more closely described by taking a $t_{2g}$ tight
binding model on the pyrochlore lattice; this model gives dispersive
subbands. Experience with Mott-localized spin-orbital systems (modelled by the
 Kugel--Khomskii hamiltonian) suggests that the spin and orbital degrees of
freedom play similar roles, and this should hold for correlated itinerant
systems as well. Indeed, the RPA treatment of the $t_{2g}$ pyrochlore
Hubbard model indicates that both spin and orbital fluctuations are
important, and that spin antiferromagnetism and orbital ferromagnetism
are competing instabilities \cite{tsunet2g}.

One way to judge the importance of correlation effects is to check to which
extent standard band structure calculations can account for the observed
properties. The LDA electronic structure of both Cd$_2$Os$_2$O$_7$ and
Cd$_2$Re$_2$O$_7$ has been determined \cite{band,harima}. It is curious that
Cd$_2$Os$_2$O$_7$ which has an unmistakable metal--insulator transition,
appears as not particularly correlated: the specific heat coefficient $\gamma$
of the metallic phase is relatively well explained by the calculated
density of states $\rho(\epsilon_{\rm F})$. In contrast, there is a
discrepancy between the LDA result and the measured $\gamma$ for
Cd$_2$Re$_2$O$_7$, and there is a mass enhancement factor
$m^*/m=\gamma_{\rm exp.}/\gamma_{\rm calc.}>1$ to account for. Actually,
the two calculations give somewhat different estimates for $m^*/m$: Singh
et al. \cite{band} imply that their $m^*/m\approx 2.4$ is not unexpected for a
superconductor, i.e., that it could be due to electron--phonon coupling,
while Harima \cite{harima} finds $m^*/m>5$, and concludes that
Cd$_2$Re$_2$O$_7$ is a strongly correlated system. Optical data indicating an
even higher $m^*/m\approx 20$ certainly add to the weight of evidence that
electron correlation is important in Cd$_2$Re$_2$O$_7$ \cite{nlwang};
however, the origin of the discrepancy between the specific heat and
optical estimates for the mass enhancement factor remains unclear.
Sakai et al. \cite{sakai} interpret susceptibility and NMR data as
indicating localized moment character, which is synonymous with
local correlations.

Taken in itself, $m^*/m$ is probably not sufficient to allow us to decide
whether either of these pyrochlore oxides is strongly correlated; neither
would such an assertion explain readily all what is seen. The large unit cell
and the threefold degeneracy of the $t_{2g}$ band are enough to show that
individual subbands have to be narrow, and therefore electron--electron
interaction is certainly important: it must be at least comparable to
the subband widths, and also to the inter-subband pseudogaps.
There is a variety of spin, and orbital, but also spin-to-orbital
correlations which may become important in the range up to room temperature.
Furthermore, the frustrated nature of the lattice precludes
any simple ordering scenario, and therefore short-range intersite, or even
multi-site, correlations have to be considered. Because of the lattice
geometry, it is likely that the short-range correlated states can be envisaged
as consisting of tetrahedron units with internal degrees of freedom. Such
basic units of the correlated electronic state may couple to the lattice,
which gives the possibility of tetramerization with the electronic degrees of
freedom still fluctuating. Alternatively, the electronic degrees of freedom
may become frozen into a collective state whose definition relies on the
unique features of the network of tetrahedra \cite{tsune,ueda}. We note that
the idea of ``bond chirality ordering'' was advocated in a recent
investigation of the low temperature lattice structure of
Cd$_2$Re$_2$O$_7$ \cite{low_T_sym}.

Here we present the results of electrical resistivity and thermopower
measurements on Cd$_2$Re$_2$O$_7$ in the pressure range 0--2 GPa.
Our basic finding is that there is a wide range of temperature (roughly
$60{\rm K}<T<200{\rm K}$) where the electronic state of the material is
``soft'', which is shown by the strong pressure dependence of the thermopower
$S$ (and to a lesser extent also of the resistivity $\rho$).
The existence, and extreme pressure sensitivity, of the so-called
second phase transition of Cd$_2$Re$_2$O$_7$ \cite{thesecond}, is only one
aspect of this behavior. We display a phase diagram based on thermopower
anomalies, including data on the pressure dependence of the second
structural transition. Measuring thermopower under pressure
is currently the only way to learn about the shape of that phase boundary.

\section{The phases of ${\boldmath{{\rm Cd}_{2}{\rm Re}_2{\rm O}_7}}$}

Cd$_2$Re$_2$O$_7$ is unique among the $5d$ pyrochlores inasmuch as it is the
only member of this group which is known to become superconducting at
$T_c= 1 - 2{\rm K}$ \cite{reprl}. However, we discuss only the
normal metallic phases, for which similarities and dissimilarities to other
systems are of interest.

The contrasting behavior of Cd$_2$Re$_2$O$_7$ and
Cd$_2$Os$_2$O$_7$ is intriguing. At room temperature $T_{\rm RT}$
both are bad metals with similar values of the nearly temperature
independent resistivity corresponding roughly to a mean free path
of the order of the lattice constant. Band structure calculation
\cite{band,harima} shows they are $5d$ semimetals with the Fermi
level position corresponding to a 1/3-, and 1/2-, filled $t_{2g}$
subband, resp. We know of no feature of the density of states, or
the Fermi surface, which would obviously account for the fact that

Cd$_2$Os$_2$O$_7$ becomes an insulator at 220K, while at almost
the same transition temperature $T_{\rm H}=$200K Cd$_2$Re$_2$O$_7$
becomes a better metal. In the notation $T_{\rm H}$, the subscript
H stands for "higher", for Cd$_2$Re$_2$O$_7$ has also a "lower"
transition temperature $T_{\rm L}$ which will be discussed later.

\begin{figure}[ht]
\centerline{\includegraphics[width=7cm]{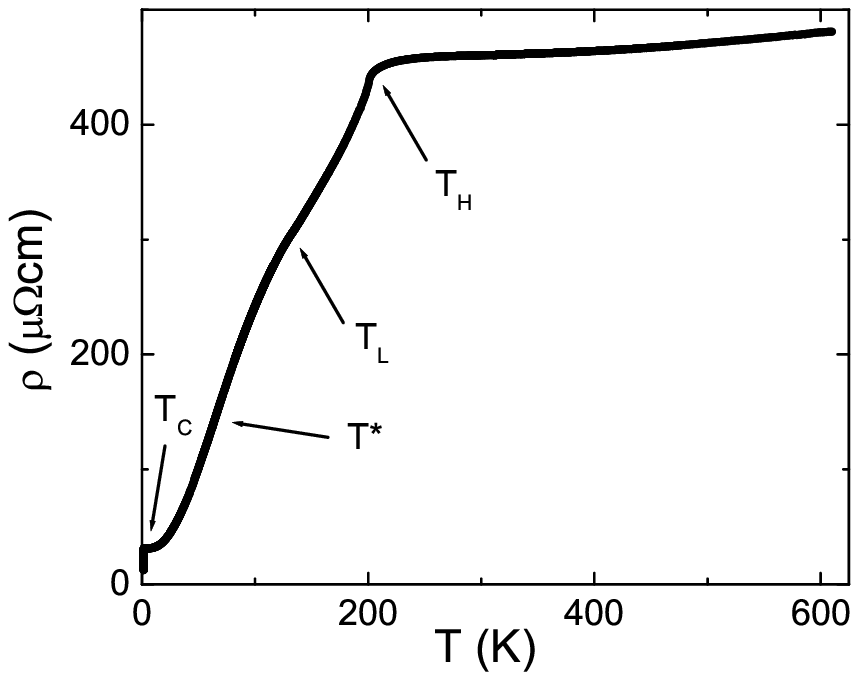}
\includegraphics[width=7cm]{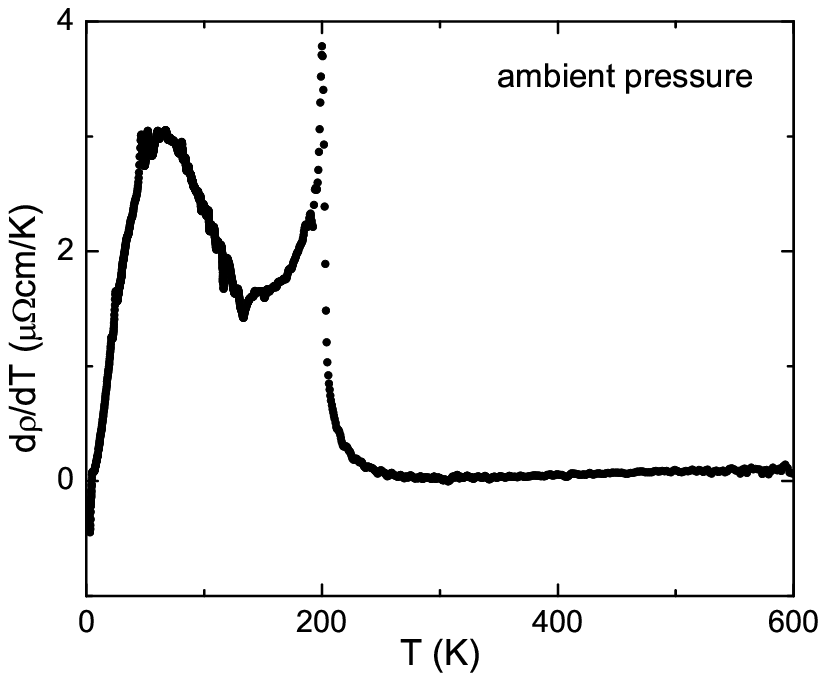}}
\caption{{\sl Left}: The temperature dependence of the resistivity
of Cd$_2$Re$_2$O$_7$ at ambient pressure. $T_{\rm H}$, $T_{\rm
L}$, and $T_{\rm c}$ signify phase transitions, while $T^*$ marks
the position of the cross-over to a low-$T$ power law regime (see
the text). {\sl Right}: The derivative of the resistivity.}
\label{fig:rho600}
\end{figure}

We illustrate the character of the phase transitions  of
Cd$_2$Os$_2$O$_7$ by the results of our measurements of the
temperature dependence of the resistivity at ambient pressure
(Fig.~\ref{fig:rho600}, left). We note that none of the previous
measurements were carried up to 600K. The extended temperature
scale makes the fundamental change in the character of
Cd$_2$Os$_2$O$_7$ at $T_{\rm H}=200$K clear. The high-$T$ phase is
a bad metal, with a resistivity which at first seems to saturate
at $\sim 500\mu\Omega{\rm cm}$, but then picks up again.

Decreasing the temperature through $T_{\rm H}=200$K, the system
becomes a good metal. The low-$T$ normal state
resistivity \cite{footnote3}  extrapolates to 30$\mu\Omega$cm. The resistivity change
around $T_{\rm H}=200{\rm K}$ could be compatible either with a
change in the scattering mechanism, or a change in carrier
concentration, or a combination of both. Whether a phase
transition occurs, has to be decided from other measurements.

Closer examination of the resistivity ($\rho$) versus $T$ plot
reveals that the slope passes through a maximum, a minimum, and
reaches a maximum again at $T_{\rm H}$, thus plotting $d\rho/dT$
vs $T$ readily offers the possibility of identifying
characteristic temperatures as either extrema, or zeroes, of the
derivative. We illustrate this with our data shown in
Fig.~\ref{fig:rho600} (right). The high-temperature peak belongs
to the bad-metal-to-good-metal transition at $T_{\rm H}$. It was
mentioned in  \cite{fluct} that the low-temperature hump at
$T^*\approx 60{\rm K}$ is also significant; however, it does not
belong to a phase transition. Rather, it is a crossover
temperature where the low-$T$ $\rho(T)$ begins to follow a
power-law behavior whose form will be discussed later.  It is
clear that other features (e.g., the minimum of $d\rho/dT$) might
have been selected. However, the status of such "characteristic
temperatures" is somewhat uncertain, and we need further evidence
to suggest that the properties of the system undergo substantial
changes at any of these points.

The continuous bad-metal-to-good-metal transition of
Cd$_2$Re$_2$O$_7$ at $T_{\rm H}=200{\rm K}$ is clearly seen in
resistivity, susceptibility, and specific heat measurements.
There is no magnetic ordering \cite{re_nmr}.  X-Ray scattering
finds new Bragg peaks, with an accompanying anomaly in the
intensity of fundamental reflections. Though at first it was
tentatively described as a cubic-to-cubic phase transition, there
is now evidence that the symmetry is lowered to tetragonal. However,
the deviation from the cubic structure is quite small,
only about 0.05\% \cite{castellan,low_T_sym}.

Evidence that a second phase transition occurs at ambient pressure
at $T_{\rm L}\approx 120{\rm K}$ ("L" stands for "lower"), was
presented by Hiroi et al \cite{thesecond}. They present a
magnified image of the $\rho$ versus $T$ plot which reveals a
minute hysteresis loop of a few K width in this region. The
transition has little effect on the resistivity, no known
signature in the susceptibility, and the specific heat
shows merely an anomaly which is two orders of
magnitude weaker than the one associated with the 200K
transition. Clearer evidence comes from magnetoresistivity measurements
which show that, after $\Delta\rho$ essentially vanishes by
the time $T$ reaches 100K, it reappears and becomes quite
anisotropic from 120K onwards. Looking at the magnetoresistivity,
it is plausible that the metallic state between 100K and 200K is
different from that below 100K. Thus Cd$_2$Re$_2$O$_7$ should
have two good metallic phases in addition to the bad metal above
$T_{\rm H}=200$K. The first thermopower data \cite{huo} are compatible with
this scenario.

Whatever the nature of the second phase transition, it is very weak.
X-Ray diffraction shows an anomaly in the temperature dependence
of the fundamental reflections, including a jump for 008 \cite{thesecond}.
A recent refinement \cite{low_T_sym} yielded the suggestive picture of
distorted tetrahedra with three unequal Re--Re distances on each triangular
plaquette. The order of the bonds allows to define a bond chirality parameter
for each triangle, and it turns out that the low-$T$ structure of
Cd$_2$Re$_2$O$_7$ can be thought of as ``ferrochiral''. Thus the 120K
transition involves a change of symmetry, but it is, strictly speaking,
not symmetry breaking: it does not belong to a symmetry lowering (i.e.,
choosing a subgroup of the original symmetry group) from the $T>120{\rm K}$
phase, but to replacing one symmetry element with another. This symmetry
characterization is compatible with the idea that the transition at
$T_{\rm L}$ is of first order.

It is not clear what really happens at either the $T_{\rm H}$ or the
$T_{\rm L}$ phase transition; in particular, whether there is an
electronic order parameter coupled to the obvious structural ones.
Though the symmetry changes are marked, we should not forget that they are
realized by quite minute \cite{footnote4}.  distortions
of the high-$T$ cubic structure; thus though both phase transitions are
literally structural transitions, it is not obvious that the structural
change is enough to explain the drastic changes in electrical properties.
In other words, we still have to search for the concomitant change of the
many-electron state, which might well be the primary phenomenon. The
X-Ray study indicates that the temperature dependence of the new
Bragg peaks is anomalously slow \cite{castellan}, which would be compatible
with the idea that the structural order parameter is secondary, induced
by some underlying electronic order parameter. Re NQR and Cd NMR
reveal that the local environment of the Re site looses trigonal symmetry at
$T_{\rm H}$, and that the character of orbital fluctuations changes at
$T_{\rm L}$ \cite{re_nmr,arai}. XPS indicates that the $t_{2g}$ bands are
split at low $T$ \cite{irizawa}. All the evidence points to significant
 rearrangement of the $t_{2g}$ subbands, with consequent changes in
their occupation, starting from $T_{\rm H}$, and continuing until
well below $T_{\rm L}$. We envisage an itinerant version of
orbital ordering transitions cite{footnote5}.

 Certainly there is ample motivation to seek a more
complete characterization of the three normal metallic phases.

Extending the measurements to higher pressures often proves enlightening.
Hiroi et al  measured the resistivity under the pressure of
1.5GPa, and at five high-pressure values between 3 and 8GPa
\cite{highpr}. 3.5GPa suffices to suppress the major structural
phase transition at $T_{\rm H}$. The weak transition at $T_{\rm L}$ seems
more sensitive to pressure. A single data point published in the less known
\cite{kotai}  indicates that $T_{\rm L}$ is suppressed to zero somewhere
beyond 2GPa. Mapping out the boundary between the two good metallic phases is
an outstanding issue.

Here we present new data about the resistivity $\rho$ and
thermopower $S$ of good-quality single crystal specimens of
Cd$_2$Re$_2$O$_7$ under pressure. For resistivity, we have more
pressure values up to 2GPa than in previous works, and at ambient
pressure, we have extended the measurement to 600K. Our
thermopower data reveal the highly anomalous nature of the "good
metallic" phases of Cd$_2$Re$_2$O$_7$. At ambient pressure, our
thermopower vs $T$ curve has a much better resolved anomaly at the
lower phase transition than the recently published \cite{huo},
allowing to identify it as a secondary minimum. We performed the
first measurements of thermopower under pressure, and present a
pressure--temperature phase diagram based on them, including new
results on $T_{\rm L}(p)$. We find a distinctive feature of the
intermediate temperature range $T^*<T<T_{\rm H}$: it is where
electrical properties are remarkably sensitive to pressure. It
stands to reason that this is a regime of continuing rearrangement
of $5d$ subbands where the electron--lattice interaction is
particularly important. In contrast, the low-temperature ($T<T^*$)
good metal, and the high-temperature ($T>T_{\rm H}$) bad metal,
are essentially pressure-insensitive.

\section{Results and Discussion}

Single crystal specimens of Cd$_2$Re$_2$O$_7$ were grown using a
vapor-transport method which is described in detail elsewhere
\cite{mandrus6}. The crystal was cut in smaller pieces of
rectangular parallelepiped shape with the dimensions
1.5x0.25x0.025mm. After placing four contacts on the sample, it
was mounted on a homemade thermopower sample holder, which fits
into a clamped pressure cell. Small metallic heaters installed at
both ends of the sample generated the temperature gradient
measured with a Chromel-Constantan differential thermocouple. The
pressure medium used in this study was kerosene, and the maximum
pressure was 2GPa. The pressure was measured using a calibrated
InSb pressure gauge.

\begin{figure}[ht]
\centerline{\includegraphics[width=7cm]{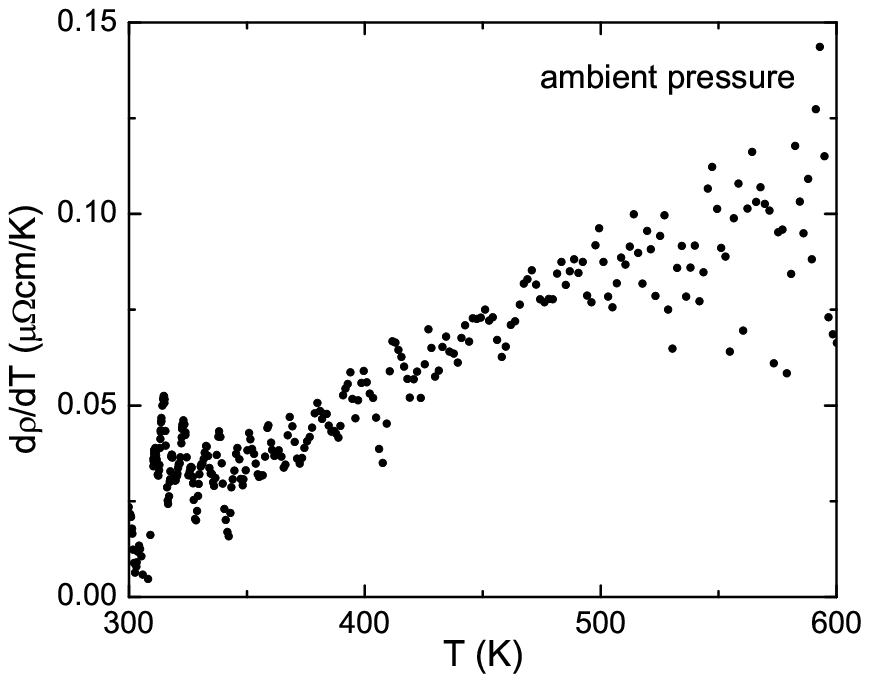}
\includegraphics[width=7cm]{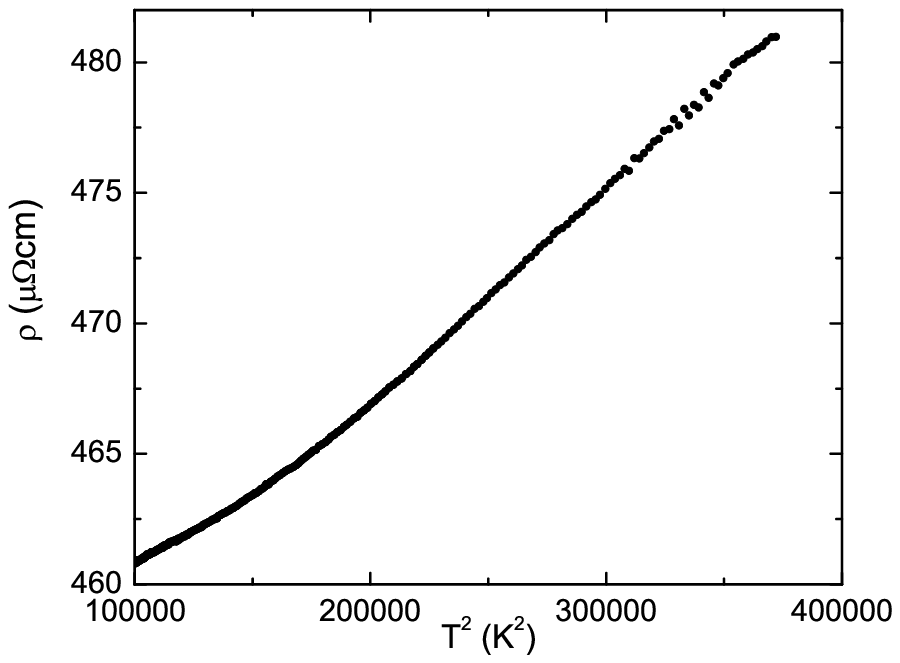}}
\caption{The high temperature resistivity. {\sl Left}: $d\rho/dT$
shows that $\rho$ is not linear in $T$. {\sl Right}: a $\rho$
versus $T^2$ fit is successful in the range 400K--600K.}
\label{fig:hightres}
\end{figure}

The temperature dependence of the ambient-pressure resistivity of
a single-crystal specimen of Cd$_2$Re$_2$O$_7$ was shown in
Fig.~\ref{fig:rho600}(left). We stress that previous measurements
did not extend up to 600K. Right above the $T>T_{\rm H}$
transition, the resistivity appears to have saturated, but the
extended scale shows up a further increase with temperature. It
strikes the eye that the high-$T$ resistivity is not linear in
$T$; this is also brought out by the enlargement of the high-$T$
part of the derivative plot (Fig.~\ref{fig:hightres}, left). For
$T\ge 400{\rm K}$, the resistivity seems to follow a $T^2$-law
(Fig.~\ref{fig:hightres}, right).  The shape of the $\rho$ vs $T$
curve deviates from what one would expect from the phonon
mechanism, and we ascribe it to inter-orbital (or inter-subband)
scattering.

We refer also to the $d\rho/dT$ derivative plot shown in
Fig.~\ref{fig:rho600} (right). The continuous metal-to-metal
transition at $T_{\rm H}=200$K is marked by a strong peak of the
derivative which is known to correlate with the susceptibility and
specific heat anomalies \cite{fluct}. The overall $T<T_{\rm H}$
behavior is similar to that known from previous measurements. The
derivative plot $d\rho/dT$ vs $T$  shows three characteristic
temperatures. The large peak on the high-$T$ side could serve to
define $T_{\rm H}$, but we prefer the definition from the
derivative of the thermopower. The broad, nearly
pressure-independent hump at $T^*\approx 60$K marks the boundary
between two regimes within the same low-temperature phase. There
is also a recognizable minimum between these two maxima; at
p=1atm, it is a rather sharp feature at $\sim$130K, and happens to
lie near the second transition temperature $T_{\rm L}$ which we
identify from thermopower data. However, we discard the
possibility of identifying $T_{\rm L}$ from this anomaly. The
reason is that a minimum in $d\rho/dT$ continues to show up at all
pressures, while there are good reasons to think that the second
phase transition is completely suppressed before $p$ reaches 1.8GPa.

The normal-state residual resistivity is 30$\mu\Omega$cm, similar
to the value given in \cite{highpr}, but about a factor of 2
higher than the value quoted in \cite{thesecond}, and in
\cite{huo}. The cooling/heating rate in our measurements was
$\approx$0.6K/min, six times faster than in the measurements of
Hiroi et al \cite{thesecond}. This might be the reason why we did
not see hysteresis in either resistivity or thermopower, though we
have a fairly dense set of data points with little scatter in the
relevant range of $T$. Nevertheless, we agree with the suggestion
that the second phase transition is weakly first order.

\begin{figure}[ht]
\centerline{\includegraphics[width=11cm]{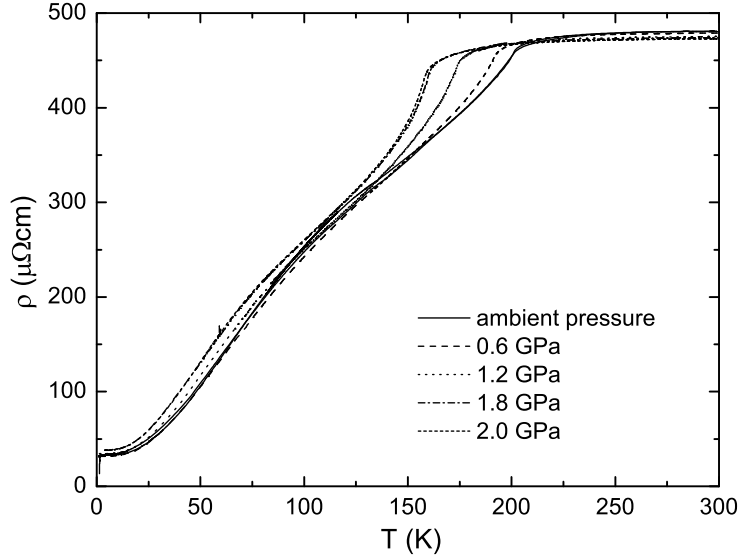}}
\caption{The temperature dependence of the resistivity of
Cd$_2$Re$_2$O$_7$ for several pressures. The sharp downturn occurs
at $T_{\rm H}$.} \label{fig:rho}
\end{figure}

\begin{figure}[ht]
\centerline{\includegraphics[width=11cm]{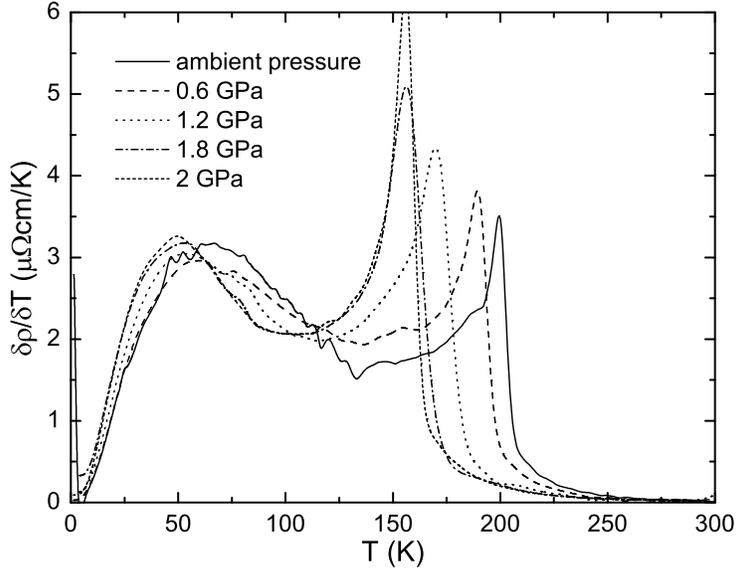}}
\caption{The temperature dependence of the derivative of the
resistivity ${\partial\rho}/{\partial T}$ for several 
pressures.}\label{fig:rhodervs}
\end{figure}

\begin{figure}[ht]
\centerline{\includegraphics[width=11cm]{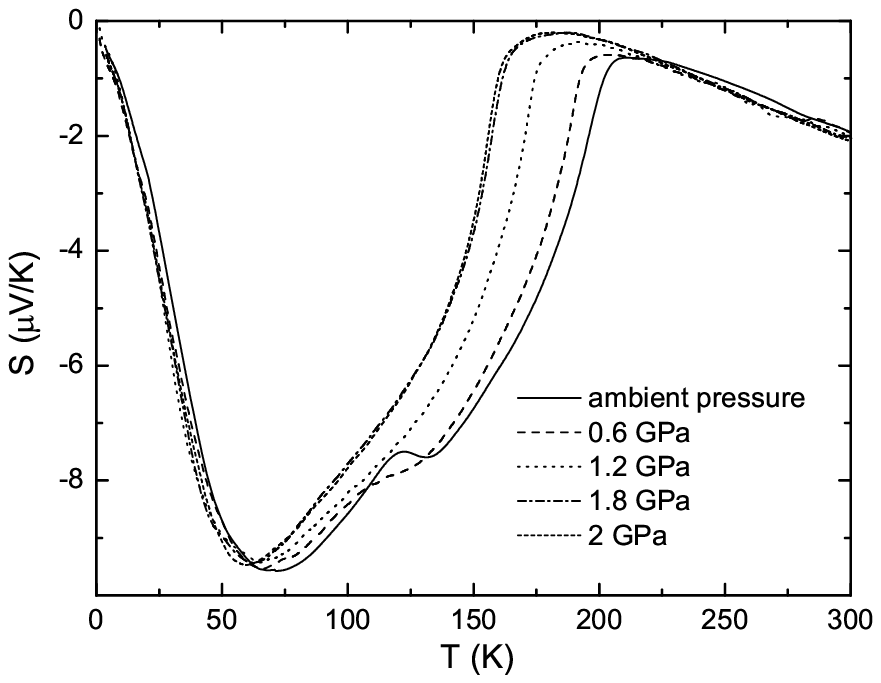}}
\caption{The temperature dependence of the thermopower $S$ for
several pressures.}\label{fig:tep}
\end{figure}

We carried out measurements at pressures of $p=1$atm, 0.6, 1.2,
1.8, and 2GPa. The temperature dependence of the resistivity
$\rho$ for different pressures (up to 300K) is  shown in Fig.
\ref{fig:rho}, the derivative curves in Fig.~\ref{fig:rhodervs},
while the thermopower $S$ is shown in Fig.~\ref{fig:tep}.

The thermopower is negative at all temperatures/pressures. The $T_{\rm H}$
phase transition has, at any pressure, an even more spectacular
signature in the thermopower than in other quantities measured
before. It is associated with a steep decrease of $S$, which
continues until at $T^*\approx 60$K $S$ reaches a minimum at which $|S|$ is
about a factor of 10 higher than at $T=T_{\rm H}$.
Below $T^*$, $S$ tends to zero in an approximately
pressure-independent manner. Our thermopower data allow
identifying three distinct regimes of temperature: (i) both $\rho$
and $S$ are essentially pressure-independent up to $T^*$; (ii):
the thermopower is quite sensitive to pressure at $T^*<T<T_{\rm H}$;
(iii): $S$ and $\rho$ are again essentially pressure-independent
in the high-temperature ($T>T_{\rm H}$) regime. Regime (ii) is the
same where X-ray studies indicate an anomalously slow
$T$-dependence of the structural order parameter. The second phase
transition, whenever found, sits in the middle of (ii).

\begin{figure}[ht]
\centerline{\includegraphics[width=7cm]{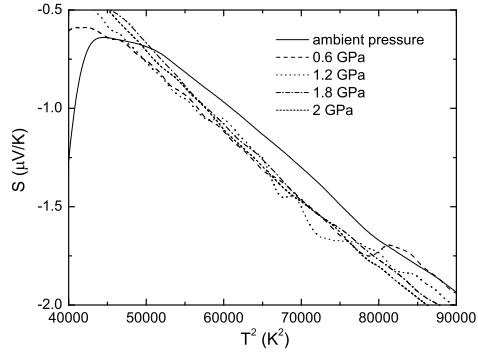}}
\caption{Thermopower vs $T^2$ in the temperature range above
$T_H$.} \label{fig:tepT2}
\end{figure}

Let us note that the overall value of the thermopower is
rather small: $|S|$ never exceeds 10$\mu$V/K. Some $5d$ elements
have higher thermopower than this, and one might have expected
that the narrower $d$-bands of our oxide give rise to a larger
thermopower. However,  $T$-dependent partial cancellation
between hole-like and electron-like contributions may explain the
observations. The degree of cancellation is smaller in the $T<T_{\rm H}$
phases; this may be compatible with the disappearance of the heavy hole
pocket found in the high-temperature (cubic phase)
band structure \cite{band}. It may also explain
why we do not find the straightforward metallic behavior $S\propto T$ at
$T>T_{\rm H}$; in fact, it is rather better described by  $S\propto T^2$ (see
Fig~\ref{fig:tepT2}).

\begin{figure}[ht]
\centerline{\includegraphics[width=11cm]{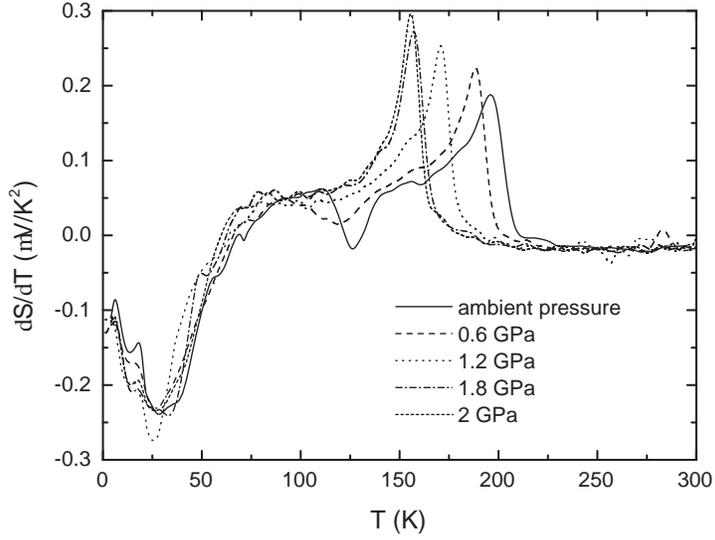}}
\caption{The temperature dependence of the derivative of the
thermopower ${\partial S}/{\partial T}$ for several
pressures.}\label{fig:tepdervs}
\end{figure}

\begin{figure}[ht]
\centerline{\includegraphics[width=7cm]{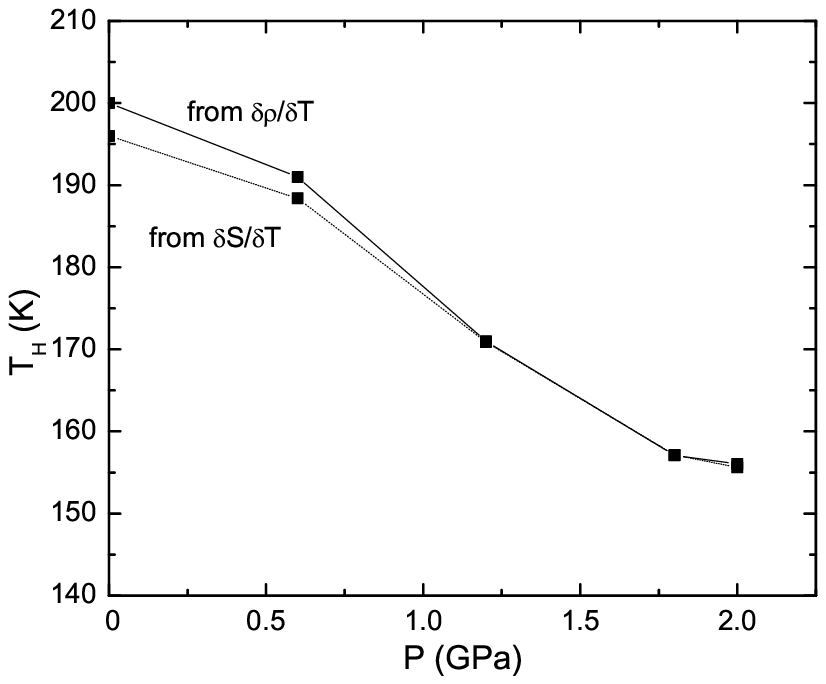}}
\caption{Using the peak positions of either $dS/dT$ or $d\rho/dT$
gives close-lying estimates for $T_{\rm H}$.}\label{fig:tccomp}
\end{figure}

Some features are more clearly seen in the derivative plot $dS/dT$
(Fig.~\ref{fig:tepdervs}). The $T_{\rm H}$ transition is well defined by
the cusp of $dS/dT$. This may seem arbitrary, but we may bring the following
argument: Loosely thinking of the thermopower as a measure of the
electronic entropy, finding an anomaly in $dS/dT$ is like finding an anomaly
in the specific heat. In any case, the peak positions of $dS/dT$ and
$d\rho/dT$ are lying pretty close (Fig.~\ref{fig:tccomp}).
It is remarkable that the $dS/dT$ peak gets {\sl sharper} under pressure. (A
similar observation was made about the $\rho$ vs $T$ curves in \cite{highpr}.)

\begin{figure}[ht]
\centerline{\includegraphics[width=7cm]{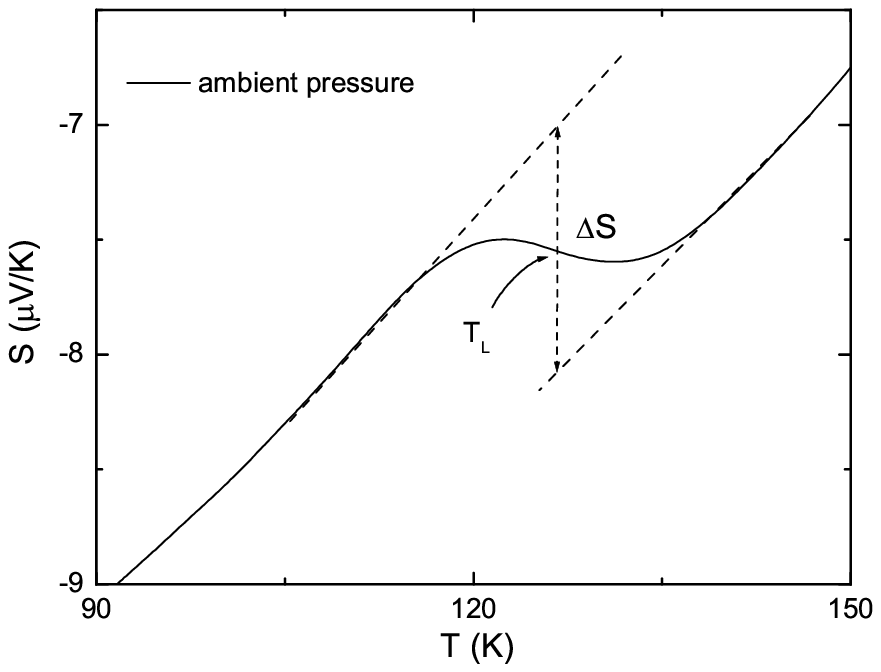}
\includegraphics[width=7cm]{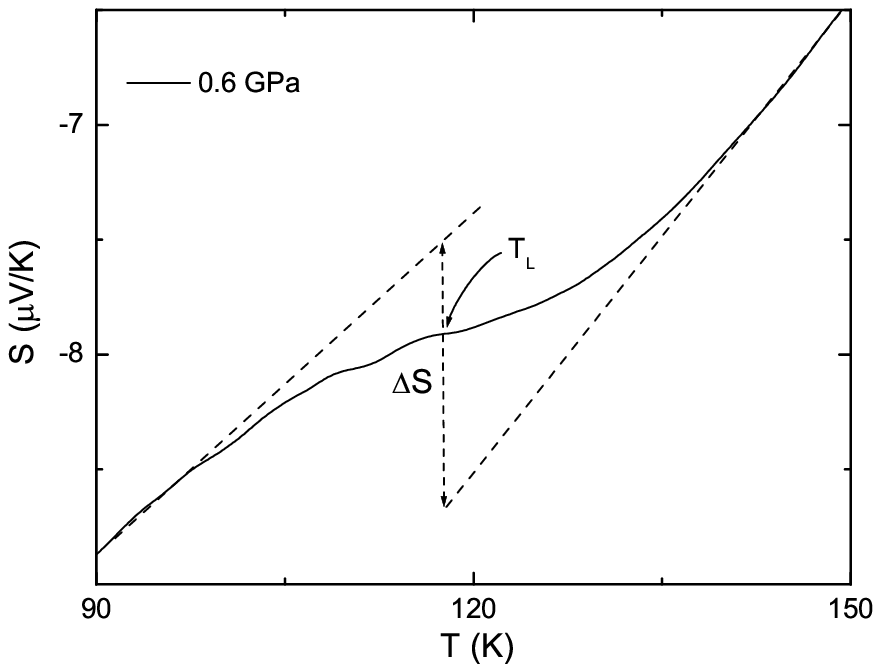}}
\caption{Tentative interpretation of the S-shaped thermopower
anomaly as a smoothed-out discontinuity for 1 bar (left), and for
0.6GPa (right).}\label{fig:tepjump}
\end{figure}

The second phase transition shows up in the $S$ vs $T$ plot
(Fig.~\ref{fig:tep}). At ambient pressure,
there is a sizeable dip in $S$ at about 120K, followed by a secondary
minimum at $\approx$130K. Let us observe that if we imagine the $S$-anomaly
sharpened (the bump and the dip getting closer, without reducing their
amplitudes), it would be consistent with a discontinuity cite{footnote6}
 of $\Delta{S}\approx -1\mu$V/K
(Fig.~\ref{fig:tepjump}, left). It suggests that we center the
phase transition at the point of inflection between the bump and
the dip, in other words at the local minimum of $dS/dT$. This
would be consistent with regarding the thermopower anomaly as the
sign of a smeared-out (and very weak) first order transition,
confirming Hiroi {\sl et al}\cite{thesecond}. We note that a
thermopower discontinuity is associated with some first order
electronic transitions, such as the valence transition of
YbInCu$_4$ \cite{ocko}.

The phase transitions shift under pressure: $T_{\rm H}$ drops to
156K under 2GPa (this is a less steep decrease than that seen in
\cite{highpr} where $T_{\rm H}(2{\rm GPa})\approx$130K. This
reflects a difference in sample quality, a point to which we
return later). We already commented on the apparently increasing
sharpness of the transition. The $T_{\rm L}$ transition gets
suppressed rather fast under pressure. At 0.6GPa the dip-and-bump
complex seems just to have shrunk to a point of inflection
(Fig.~\ref{fig:tepjump}, right): the local minimum of $dS/dT$
reaches 0 at 115K (Fig.~\ref{fig:tepdervs}). At 1.2GPa there is
not even a point of inflection, but one may still risk identifying
the bottom of a valley in $dS/dT$ at $\approx$105K. The
thermopower curves are perfectly smooth in the region
$T^*<T<T_{\rm H}$ at higher pressures, thus the second phase
transition certainly vanishes somewhere below 1.8GPa.

\begin{figure}[ht]
\centerline{\includegraphics[width=11cm]{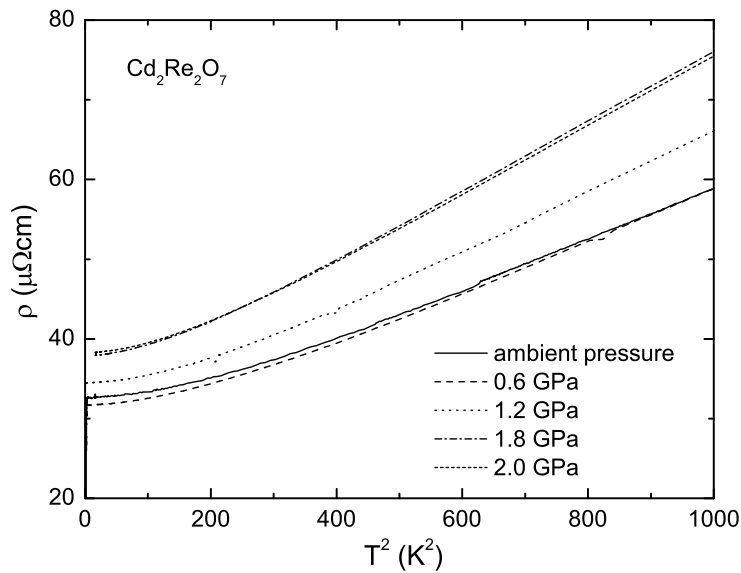}}
\caption{The resistivity in the range 1--30K, plotted as a
function of $T^2$.}\label{fig:rhoT2}
\end{figure}

\begin{figure}[ht]
\centerline{\includegraphics[width=11cm]{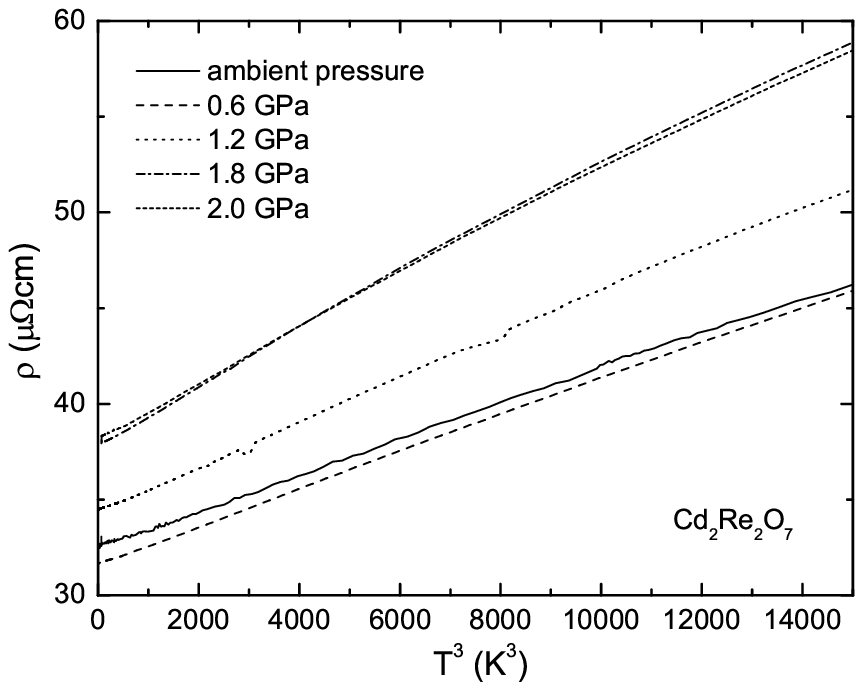}}
\caption{The resistivity in the range 1--25K, plotted as a
function of $T^3$.}\label{fig:rhoT3}
\end{figure}

\begin{figure}[ht]
\centerline{\includegraphics[width=11cm]{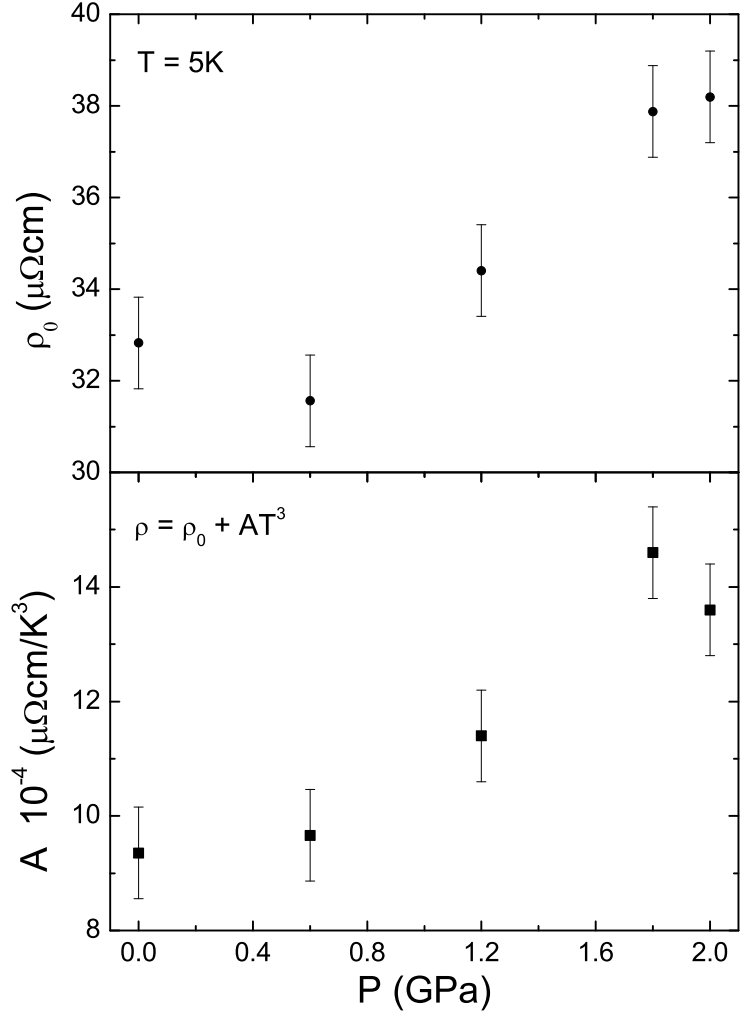}}
\caption{The pressure dependence of $\rho_0$ and
$A_3$.}\label{fig:A3}
\end{figure}

\begin{figure}[ht]
\centerline{\includegraphics[width=11cm]{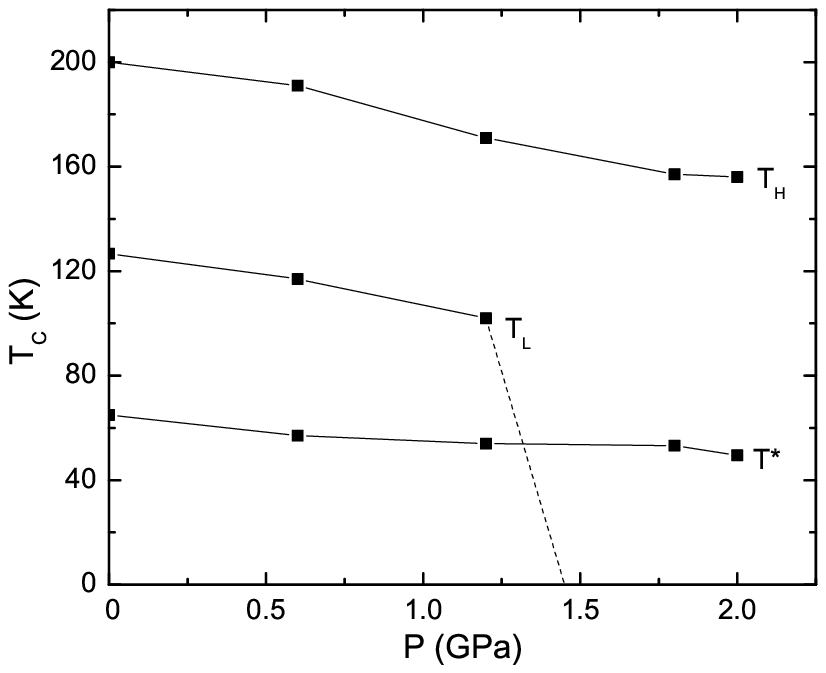}}
\caption{The $T$--$p$ phase diagram derived from thermopower and
resistivity measurements. Definitions: $T_{\rm H}$ from the peak
in $dS/dT$; $T_{\rm L}$ from the minimum of $dS/dT$; $T^*$ is the
position of the low-$T$ broad maximum of $\rho$.}\label{fig:phase}
\end{figure}

Now we return to our resistivity data.  The $T>T_{\rm H}$ resistivity
is large, essentially pressure-independent, and increases slowly
with $T$, indicating the presence of a strong scattering mechanism
specific to the pyrochlore structure. The high-temperature phase
is cubic, thus the restoration of the orbital degree of freedom gives an extra
scattering mechanism \cite{footnote7}.

The observed lack of a magnetic field dependence of the thermopower
also indicates that spin disorder scattering is
not important in Cd$_2$Re$_2$O$_7$ \cite{huo}.

Below $T_{\rm H}$ the resistivity begins to decrease sharply,
following a roughly linear $T$-dependence down to about 50K,
below which it bends over into a seemingly power-law regime which
has been fitted either with $T^2$ or with $T^3$ dependences
\cite{reprl,highpr,huo}. $\rho=\rho_0+A_2^*T^2$ with an enhanced value
$A_2^*/A_2\propto (m^*/m)^2$ would be considered
typical of strongly correlated systems, and it would be expected to show up
here, since the specific heat shows enhancement. However, we found
that the $T^2$ law would give an acceptable fit only in a narrow
range of $T$, and even then at low pressures only (Fig.~\ref{fig:rhoT2}).
In agreement with previous work \cite{reprl}, we find that a
$\rho=\rho_0+A_3T^3$ fit works better (Fig.~\ref{fig:rhoT3}).
The $T^3$-law is to be ascribed to electron--phonon rather than to
electron--electron scattering. We did not attempt to fit with a combination of
$T^2$- and $T^3$-like terms. Undoubtedly one would find a $T^2$ contribution,
but certainly not so large as to satisfy the Kadowaki--Woods relation.

The pressure dependence of $\rho_0$ and $A_3$ is given in
Fig.~\ref{fig:A3}. A moderate increase in $A$ is compatible with
increased electron--phonon interaction in a compressed lattice. We
call attention to the fact that the pressure dependence of
$\rho_0$ is quite weak, even at 2GPa we find less than
40$\mu\Omega$cm. This is to be contrasted with the results of
Hiroi et al. who find an about sixfold increase in $\rho_0$ at
1.5GPa, and a twofold increase in $A_2^*$ \cite{highpr}. The
authors of Ref.\cite{highpr} would agree that correlation, as
manifested in $A_2^*$, is weak at ambient pressure, but argue that
its role is increasing at higher pressures. Our pressure range is
limited to 2GPa, but within this range, we do not detect a
tendency towards heavy fermion behavior. Obviously, there is
sample dependence in the observed behavior; we think that the weak
pressure dependence of $\rho_0$
 is an indication of the good quality of our samples.

There is apparent difficulty in deciding what happens to the
electronic structure below 200K. The difficulty may arise partly
from trying to reduce everything to a single number, the
enhancement factor of the effective density of states
$z^{-1}={\cal N}^*(\epsilon_{\rm F})/{\cal N}(\epsilon_{\rm F})$,
which would then appear also as the enhancement of the thermal
effective mass $z^{-1}=m^*/m=\gamma^*/\gamma$, and similarly also
in the spin susceptibility, and the plasma frequency
\cite{nlwang}. Much of the previous discussion relied on the idea
that the structural transitions open pseudogaps in the spaghetti
of $t_{2g}$ subbands, and consequently ${\cal N}^*(\epsilon_{\rm F})$ 
is reduced \cite{fluct}. This seems to agree with the
reduction in susceptibility \cite{sakai}, but makes it difficult
to understand why the conductivity is substantially increased. One
can argue \cite{fluct} that heavy carriers got eliminated, which
at $T>200{\rm K}$ did not contribute to conductivity but rather
acted as scatterers. Removing narrow subbands from the vicinity of
$\epsilon_{\rm F}$ would mean that ${\cal N}(\epsilon_{\rm F})$ is
reduced. The temperature dependence of $\chi$, the Knight shift
and $(TT_1)^{-1}$ measured by $^{111}$Cd NMR are well described by
assuming a reduced ${\cal N}(\epsilon_{\rm F})$ \cite{re_nmr}.
However, high-resolution photoemission data were interpreted in
terms of a low-$T$ enhancement of ${\cal N}(\epsilon_{\rm F})$
\cite{irizawa}.

Our thermopower measurements do not allow to infer 
${\cal N}(\epsilon_{\rm F})$. Using the simplest picture of a correlated
one-band model \cite{merino} the thermopower 
$S\propto -(k_{\rm B}/|e|) z^{-1} ({\cal N}^{\prime}(\epsilon_{\rm F})
/{\cal N}(\epsilon_{\rm F}))$ measures the band asymmetry about
$\epsilon_{\rm F}$ rather than the density of states. Our data are
thus indicative of a strongly increasing asymmetry below 200K. We
may envisage a heavy subband gradually crossing out from the
vicinity of $\epsilon_{\rm F}$, which gives a strong contribution
to ${\cal N}^{\prime}(\epsilon_{\rm F})$.

A pressure--temperature phase diagram based on plotting the
characteristic temperatures $T_{\rm H}$, $T_{\rm L}$, and $T^*$ is shown in
Fig.~\ref{fig:phase}. The last data point for $T_{\rm L}$ at 1.2GPa
is, as we have seen, rather tentavive; in any case, $T_{\rm L}$
is suppressed fast. This is in general agreement with a previous
observation \cite{kotai}, but our critical pressure for $T_{\rm L}$
is rather lower. In contrast, our $T_{\rm H}$ vs pressure curve would
lie above that found by Hiroi et al \cite{highpr}. 
$T^*$ is always defined by the peak position of the broad 
maximum of $d\rho/dT$. It is rather pressure-independent,
and approximately coincides with the low-$T$ minimum of $S$.
We do not think it associated with a phase transition but it is
nevertheless significant since it marks the crossover
from a stable good metallic state into a fluctuating intermediate-$T$
state. We may think of it as the coherence temperature of the low-$T$
electronic structure.

Since we did not find the third normal metallic phase at 1.8GPa, we drew a
dashed line to show that $T_{\rm L}$ should drop to zero somewhere between 1.2
and 1.8 GPa. In principle, a line of first-order transitions could
terminate at a critical point at some finite temperature, but
the recent finding of a symmetry change at $T_{\rm L}$ \cite{low_T_sym}
rules this out: the phase boundary has to continue down to $T=0$.

One may ask what our final answer is to the question whether
Cd$_2$Re$_2$O$_7$ is a correlated electron system. Recalling the conflicting
inferred results for
${\cal N}^*(\epsilon_{\rm F})/{\cal N}(\epsilon_{\rm F})$,
we note that seeking the same enhancement factor
in various quantities originates from the underlying picture of a one-band
model. Having many kinds of spin and orbital correlations (or, alternatively,
intra- and inter-subband correlations) should allow a
contradiction-free interpretation. The evidence from specific heat alone
should suffice to say that  Cd$_2$Re$_2$O$_7$ is a correlated system. The
difficulty
that the susceptibility at the same time decreases, may be resolved if we
postulate that singlet intersite spin correlations can be induced by
electron--electron interaction in a frustrated itinerant system (a similar
result was found for the trellis lattice \cite{fluct_exch}).

\section{Conclusion}

We presented resistivity and thermopower measurements on single
crystal samples of Cd$_2$Re$_2$O$_7$ under five different pressure values
up to 2GPa, and used the data to analyze the phase diagram in the $T$--$p$
plane. The pressure dependence of the thermopower shows that the wide
temperature interval between the upper structural transition (at ambient
pressure $T_{\rm H}=200$K) and the coherence temperature
$T^*\approx 60$K is a regime of continuing rearrangement of the
electronic structure. The cubic-to-tetragonal transition at 200K does not
immediately lead to a stable low-temperature phase, but rather to a state
with electronic and structural ambiguity. A particular manifestation of this
behavior is the appearance of a second structural phase transition
which, is however, confined to relatively low pressures ($<1.8$GPa),
while the overall features of the system remain the same at higher pressures.

Our results indicate that both coupling to the lattice, and the strongly
temperature dependent redistribution of the electrons over the $t_{2g}$
orbital states (or alternatively the $t_{2g}$ subbands) are important
for understanding the behavior of this frustrated itinerant system.

\begin{acknowledgements}

P.F. is grateful to H. Harima, M. Takigawa, K. Ueda, and O.
Vyaselev for enlightening discussions, and acknowledges support by
the Hungarian grants OTKA T 038162, OTKA T 037451, and AKP
2000-123 2,2. Work at the University of Tennessee is supported by
NSF DMR-0072998. Oak Ridge National Laboratory is managed by
UT-Battelle, LLC, for the US Department of Energy under contract
DE-AC05-00OR22725. The work in Lausanne was supported by the Swiss
National Science Foundation and its NCCR Network "Materials with
Novel Electronic Properties".

\end{acknowledgements}


\begin{thebibliography}{00}
\bibitem{footnote1} In the general formula A$_2$B$_2$O$_7$, both the A and B
sublattices can be envisaged as a network of corner sharing  
tetrahedra, which is nowadays referred to as the pyrochlore lattice.
\bibitem{footnote2} The isotropic Heisenberg model with
nearest-neighbour interactions has no ordering transition on the
pyrochlore lattice, either for S=1/2 \cite{canals}, or for
classical spins \cite{chalker}. The spin liquid phase may be
unstable against small perturbations like second neighbour
coupling, in which case the question of the kind of spin ordering
pattern arises \cite{pinettes}. However, it is difficult to see
how a weak effect of this kind could cause a robust
metal--insulator transition at 220K.
\bibitem{footnote3} Fig.~\ref{fig:rho600} also shows the
superconducting transition at $T_c\approx 1{\rm K}$. However, this
will not be discussed in this paper, since we could not
systematically investigate the superconducting phase under pressure.
\bibitem{footnote4} Strictly speaking, we know only that the
  overall changes in unit cell dimensions are very small. It may still turn out
   that individual atomic motions within the unit cell
  are relatively large, which would allow that the phase transitions are
  primarily structurally driven (band--Jahn--Teller transitions).
  However, we take a different view.
\bibitem{footnote5} Since the lattice responds
to the change in the electronic state, the transitions may be
called band--Jahn--Teller transitions. When speaking rather of
orbital order, we imply that the phenomenon is primarily driven by
electron--electron interactions, and that it would happen even if
we held the lattice fixed. A future calculation has to decide if
this scenario is correct.
\bibitem{footnote7} The high-$T$ behavior of Cd$_2$Os$_2$O$_7$ is
pretty much the same, but Os$^{5+}=5d^3$ would not have an orbital
degree of freedom in the high-spin state. However, allowing for the presence
of the low-spin configuration gives orbital degeneracy, and allows the two
systems behave similarly.
\end{thebibliography}
\end{document}